# Revealing Mammographic Phenotypes in Deep Learning Breast Cancer Risk Models

**Ruiyu Jia**[1] RJ2635@NYU.EDU
**Yanqi Xu** [2] YX2105@NYU.EDU
**Yuxuan Chen**[4] YC7087@NYU.EDU
**Yiqiu Shen**[2,3] YS1001@NYU.EDU
**Laura Heacock**[3] LAURA.HEACOCK@NYULANGONE.ORG

[1] Department of Mathematics, New York University Shanghai, China
[2] Center for Data Science, New York University, USA
[3] Department of Radiology, NYU Langone Health, USA
[4] Perlmutter Cancer Center, NYU Langone Health, New York, USA

## Abstract

Mammogram-based deep learning models have improved breast cancer risk prediction, but the learned imaging patterns remain underexplored. Existing interpretability methods rely on single-image saliency maps, failing to identify recurring mammographic phenotypes across large patient cohorts. By clustering patch embeddings from a pre-trained model, Mirai, we isolate recurring phenotypes linked to 5-year cancer risk. Analyses show risk-increasing phenotypes capture complex structures (e.g., dense tissue, microcalcifications) and shortcut artifacts (e.g., clips). These phenotypes correlate strongly with older age and higher BI-RADS density. Our framework connects tissue patterns to AI risk scores, revealing clinical signatures and potential latent model confounders.

## 1. Introduction

Breast cancer risk prediction estimates the chance that a currently healthy individual will develop breast cancer in the future. This differs from cancer diagnosis, which aims to detect tumors that already exist. Deep learning models, such as Mirai (Yala et al., 2021), have improved risk prediction by analyzing breast imaging, primarily mammograms (Xu et al., 2026) and shows that mammograms contain prognostic signals of future cancer risk (Lehman et al., 2022). However, existing models often operate as black boxes, leaving the learned imaging patterns unknown. Existing interpretability tools, such as saliency maps, only highlight important regions in individual images. They cannot identify or analyze recurring mammographic phenotypes at population level (Pertuz et al., 2023; Shifa et al., 2025; Gurmessa et al., 2024).

To address this gap, we present an interpretation pipeline (Figure 1) that reveals the mammographic phenotypes deemed important by trained deep learning risk models. Unlike methods that explain single images in isolation, our approach systematically mines globally shared imaging patterns and uses statistical tools to reveal the association between these phenotypes and predicted cancer risk. We demonstrate this framework using Mirai, a validated deep learning model for 1- to 5-year breast cancer risk prediction (Yala et al., 2022). Our pipeline is model-agnostic and can easily be applied to other image-based risk models.

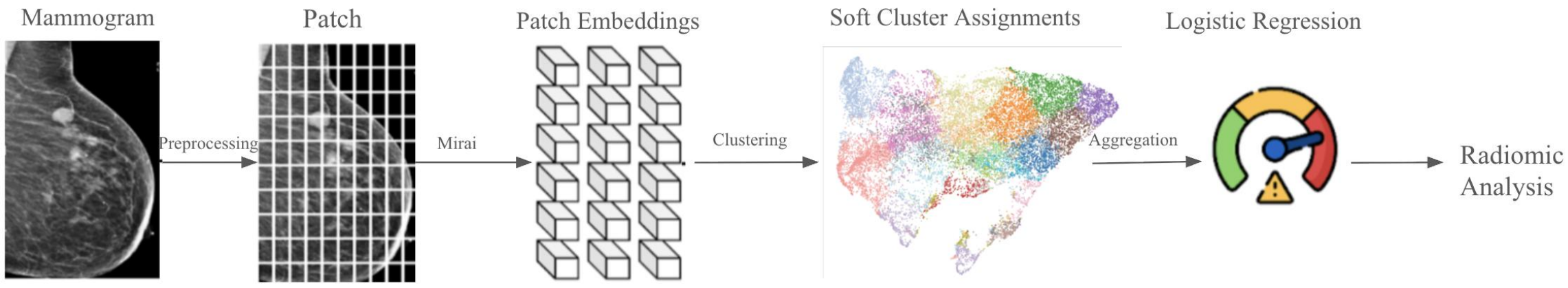

Figure 1: Overview of the proposed pipeline.

## 2. Methods and Dataset

**Dataset.** We used a screening full-field digital mammography dataset collected at NYU Langone Health. It contains 72268 mammograms from 47174 patients. Each exam included the four standard views, which were resized to 1664 × 2048 pixels. We split the dataset at the patient level into clustering (35206 exams), validation (10972 exams), and test (26090 exams) cohorts. We excluded cases where cancer is visualized from the validation and test cohorts. Cancer outcome were confirmed using pathology.

**Patch-Level Embedding and Clustering.** We cropped each mammogram into 224 ×224 patches, discarding those with over 90% background pixels. We encoded the remaining patches using the pre-trained Mirai image encoder and applied soft k-means clustering to the resulting embeddings. Each cluster represents a phenotype. To capture the overall composition of the breast, we summed and normalized these patch-level assignments: each dimension of this vector indicates the proportion of the image's tissue area associated with a particular cluster. We then fit a logistic regression model on these image-level vectors to predict the patient's 5-year cancer risk. The number of clusters a hyperparameter tuned on the validation set.

**Statistical Analysis.** Based on the regression coefficients, we identified phenotypes posi- tively and negatively associated with cancer risk. To characterize these phenotypes quantita- tively, we extracted 93 IBSI-compliant radiomic features for each patch using PyRadiomics (van Griethuysen et al., 2017; Zwanenburg et al., 2020). For each radiomic feature, we compared its distribution between the positive and negative clusters using Mann-Whitney U tests, quantifying effect sizes with Cohen's d. To assess associations with established clinical risk factors, we analyzed how the proportion of patches assigned to positive phe- notypes varied across patient subgroups. We used a Kruskal-Wallis test to compare this proportion across BI-RADS density categories, and a Mann-Whitney U test to compare the ages of patients whose mammograms were predominantly composed of positive versus negative phenotypes (Boyd et al., 2007).

## 3. Results and Discussion

Radiomic comparison between positive and negative clusters revealed systematic differences, with the 20 largest-effect features shown in Figure 2(a). Compared to negative phenotypes, positive phenotypes (Cohen's d = 0.41–0.56) showed higher Contrast, Complexity, Difference Average, Joint Entropy, and Small Area Emphasis, whereas negative phenotypes showed higher Id/Idm and Inverse Variance. These findings indicate that positive patches are locally more heterogeneous in their tissue pattern, while negative patches are more homogeneous and more likely to be radiolucent. Visual review of the highest-weight positive clusters suggests that Mirai encoder captures both biologically plausible tissue phenotypes and non-biologic shortcut signals Figure 2(b)). Representative patches included dense focal fibroglandular tissue (C37, C65), linear stromal patterns with higher Cluster Shade or Joint Entropy consistent with architectural-distortion-like complexity (C19, C116), scattered microcalcifications without a focal suspicious grouping (C126, C63; red boxes), and clusters dominated by biopsy clip markers or skin markers (C70, C74), which may also serve as proxies for prior clinical history associated with elevated future cancer risk.

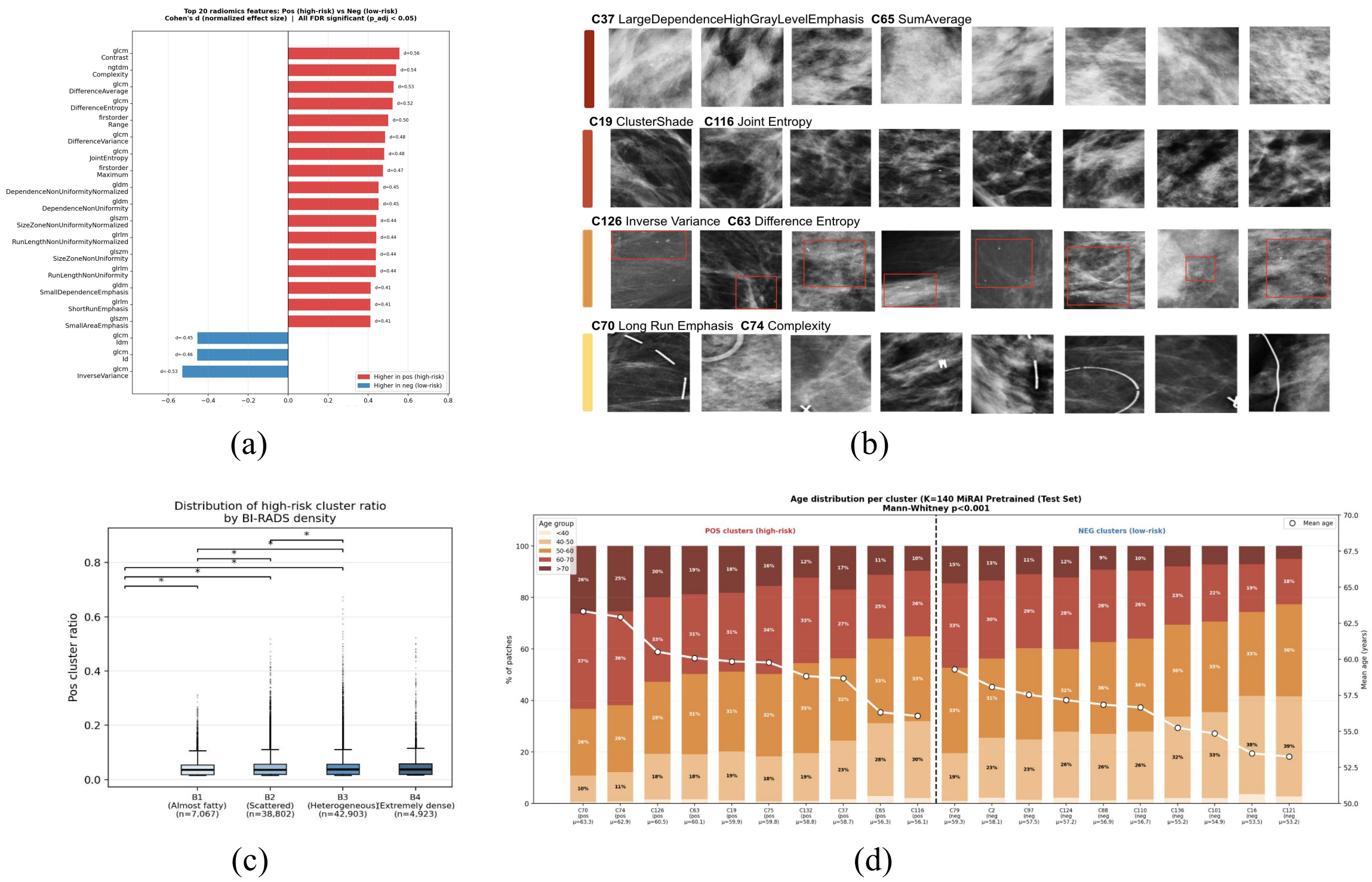

Figure 2: (a) Top 20 radiomic features distinguishing positive and negative clusters. (b) Representative patches from high-weight positive clusters. (c) Image-level proportion of positive patches across BI-RADS density categories. (d) Age distribution for patients contributing to positive and negative clusters.

Association analyses indicate that some learned phenotypes act as proxies for estab- lished risk factors. The image-level proportion of positive patches increased monotonically across BI-RADS density categories (Kruskal–Wallis $p < 0.001$; Figure 2(c)), and patients contributing to positive clusters were significantly older than those contributing to negative clusters (mean age 59.2 vs. 56.3 years; Mann–Whitney $p < 0.001$; Figure 2(d)). Notably, while fibroglandular density is a known long-term risk factor, our results demonstrates that localized patch-level stromal and tissue patterns provide risk-associated information beyond whole-breast density. Overall, the proposed framework offers a scalable and interpretable approach for linking AI model's risk prediction to localized mammographic phenotypes.